\documentclass[secnumarabic, twocolumn, amssymb, nobibnotes, apl]{revtex4}

\usepackage{graphicx}
\usepackage{pslatex}
\usepackage{xspace}
\usepackage{subfigure}

\usepackage{booktabs}

\newcommand{\bfig}{\begin{figure}}
\newcommand{\efig}{\end{figure}}
\newcommand{\bit}{\begin{itemize}}
\newcommand{\eit}{\end{itemize}}

\begin{document}

\preprint{}

\title{Bipolar Charge Transport in Poly(3-hexyl thiophene)/Methanofullerene Blends:\\A Ratio Dependent Study}

\author{A.~Baumann$^1$}\email{abaumann@physik.uni-wuerzburg.de}

\author{J.~Lorrmann$^1$}

\author{C.~Deibel$^1$}\email{deibel@physik.uni-wuerzburg.de}

\author{V.~Dyakonov$^{1,2}$}
\affiliation{$^1$ Experimental Physics VI, Julius-Maximilians-University of W\"urzburg, 97074 W\"urzburg, Germany}
\affiliation{$^2$ Bavarian Center for Applied Energy Research e.V. (ZAE Bayern), 97074 W\"urzburg, Germany}

\date{\today}

\begin{abstract}
We investigated the charge carrier mobility in pristine poly(3-hexyl thiophene-2,5-diyl) (P3HT):[6,6]-phenyl-C$_{61}$ butyric acid methyl ester (PCBM) blend devices by applying the time resolved photoconductivity experiment in dependence on the donor:acceptor ratio. We observe a bipolar transport in all studied samples ranging from pure polymer to polymer:fullerene with 90\% PCBM content. For the ratios P3HT:PCBM 1:4 and 1:1 we observe two transit times in the electron current transients, as well as hole double transients for P3HT:PCBM 1:2. We find high hole and electron mobilities in the order of $10^{-3}$--$10^{-2}$ cm$^2$/Vs for a concentration of 90\% PCBM in the blend. 
\end{abstract}


\maketitle

Due to its promising potential of low cost fabrication, organic photovoltaics has attracted a lot of attention during the last decade. Bulk heterojunction (BHJ) solar cells based on blends of poly(3-hexyl thiophene) P3HT and phenyl-C$_{61}$ butyric acid methyl ester PCBM belong to the most promising candidates for this application, as efficiencies in the range of 5\% have been achieved recently~\cite{green2008}.\\

The mobility of photogenerated charges is one of the crucial factors determining  the performance of organic solar cells~\cite{deibel2008a}, its impact being more complex than following any ``faster equals better'' stereotype. Nevertheless, one of the prerequisites for a well-performing BHJ solar cell is that the respective mobilities for transporting electrons on the fullerene and holes on the conjugated polymer are rather balanced, in order to avoid a space charge building up. However, as the blend of two different material types cannot be described by a simple superposition of the single material's properties, a prediction of the bipolar charge transport in a solar cell is not straightforward. Tuladhar et al.~\cite{tuladhar2005} investigated the dependence of the charge carrier mobility in poly[2-methoxy-5-(3',7'-dimethyloctyloxy)-1,4-phenylenevinylene (MDMO-PPV) with varying PCBM content using time-of-flight transient photoconductivity (TOF). They found an increase of the mobility for both carrier types with an increasing fullerene fraction. Additionally, ambipolar transport was found in PCBM embedded in an insulating polystyrene matrix. For the nowadays more relevant combination of P3HT:PCBM, field effect transistor (FET) experiments were reported in literature, observing an ambipolar transport in a limited range of blend ratios~\cite{shibao2007,vonhauff2006a}. However, field effect measurements consider neither photogenerated charges nor low carrier concentrations. 
There is only one publication studying P3HT:PCBM blends with varying donor:acceptor ratio using the relevant TOF technique. Huang et al.~\cite{huang2005} found an anomalous transition from dispersive to nondispersive and back to dispersive transport for an increasing PCBM fraction. The authors use films around one micrometer thickness, and do not comment on the observed dispersivity which might be influenced by the layer thickness~\cite{borsenberger1992}. In this context, the ratio dependent electron and hole mobilities would have been of particular interest, but are not reported.
In this letter, we aim at addressing the lack of available experimental data on electron and hole mobilities in P3HT:PCBM blends with varying ratio under controlled conditions.
We determined the mobility for both carrier types by the TOF method in the regimes of non-dispersivity and low carrier concentration. For different electric fields, we observe a bipolar transport over the whole range of PCBM fractions in a P3HT matrix. Furthermore, at several P3HT:PCBM ratios we observe double transients, corresponding to two different field dependent carrier mobilities. We discuss the potential origin of these findings.\\

Bulk heterojunction solar cells were prepared by slow drying a mixture of P3HT:PCBM made from solution in chlorobenzene on poly(3,4-Ethylendioxythiophen):polystyrolsulfonate covered indium tin oxide (ITO)/glass substrates. A semitransparent aluminum anode (thickness $d\sim20nm$) was evaporated thermally. P3HT was purchased from Rieke Metals, PCBM from Solenne. All materials were used without further purification. The active layer thickness ranges from approximately 2 $\mu$m to 9 $\mu$m, depending on the blend ratio and the concentration of the chlorobenzene solution, as monitored by a Dektak profilometer. An overview of the investigated samples is given in Table~\ref{tab:samples}.
\begin{table}[htb]
	\centering
	\caption{The P3HT:PCBM (D:A) ratio, the PCBM content $x$, the concentration $c$ of the used chlorobenzene solution, the determined thickness $L$ of the investigated samples, the absorption coefficient $\alpha$ and the used laser wavelength $\lambda$ is shown}
	\small
	\begin{tabular}{c*{6}{@{\extracolsep{6mm}}c}}\toprule\toprule
	D:A & $x$ & $c$ & $L$ & $\alpha$ & $\lambda$ \\ 
	&  & g/l  & $\mu$m & m$^{-1}$ & nm\\ \midrule
	1:0 & 0 & 10 & 2.2 & $1.3\times10^7$ & 500\\
	10:1 & 0.09 & 20 & 8.5 & $9.4\times10^6$ & 500 \\ 
	4:1 & 0.2 & 20 & 4.8 & $8.4\times10^6$ & 500  \\  
	2:1 & 0.33 & 20 & 5.6 & $7.2\times10^6$ & 500 \\
	1:1 & 0.5 & 20 & 4.2 & $7.7\times10^6$ & 500 \\
	1:2 & 0.66 & 30 & 6.0 & $7.0\times10^6$ & 335\\  
	1:4 & 0.8 & 30 & 4.3 & $7.7\times10^6$ & 335 \\ 
	1:10 & 0.9 & 30 & 6.5 & $8.6\times10^6$ & 335 \\ \bottomrule\bottomrule
	\end{tabular}
	\label{tab:samples}
\end{table}
The TOF experiments were performed at different electric fields in the range of about $3\cdot 10^{6}$ V/m to $4\cdot 10^{8}$ V/m. Carriers were generated by a short 5 ns nitrogen laser pulse equipped with a dye unit ranging from 335 nm to 500 nm in order to ensure sufficient absorption, as we found that samples with more than 50\% PCBM show a blue shift of the maximum absorption. The penetration depth of the laser was less than 10\% of the sample thickness. An external electric field is applied by an Agilent 33250A waveform generator and a FLC Electronics A600 voltage amplifier, which separates the photogenerated charge pairs and, depending on the field direction, transports electrons (holes) from the ITO side (Al electrode) through the device active layer to the Al cathode (ITO anode). The generated transient current is amplified by a FEMTO DHPVA voltage amplifier and detected by a Tektronix digital oscilloscope. The mobility can then be calculated from the transit time of the carriers, the sample thickness $d$ and the electric field~\cite{scher1975}. We calculate the quantum yield for the generation of electrons and holes in our experiment. With an energy of around 30 $\mu$J/cm$^2$ on the sample, including losses at Al and ITO electrode, we estimate a number of incident photons of about 7$\times10^{17}$m$^{-2}$. On the other hand, the number of extracted charges can be calculated to be around 6$\times10^{16}$m$^{-2}$, which is comparable to concentrations existing in related experiments on the studied material system, such as Photo-CELIV measurements at similar laser intensities~\cite{deibel2008b}. Thus, depending on the internal quantum efficiency, beween 10 and 50\% of the generated charges are extracted in the TOF experiment within the time frame of our measurements.
All experiments were performed at room temperature.\\
\begin{figure}
	\includegraphics{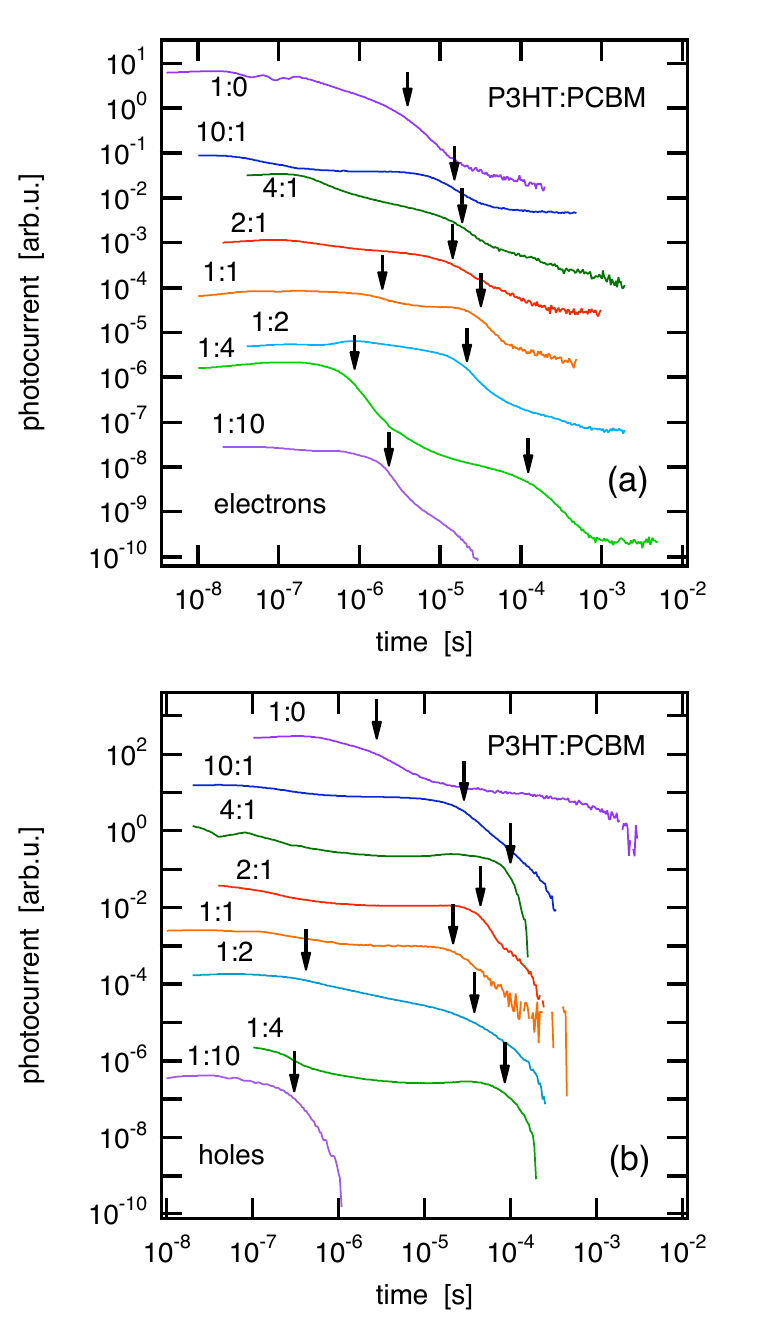}
	\caption{(Color online) (a) Electron and (b) hole current transients for the different blend ratios at electric fields from about $3\cdot 10^{6}$ V/m to $4\cdot 10^{8}$ V/m; the arrows in the graph indicate the field dependent transit times.}
	\label{fig:Fig1}
\end{figure}

Fig.~\ref{fig:Fig1} shows the current transients of electrons and holes for the studied blend ratios at different electric fields. The arrows in the graphs indicate the observed transit time in the measured transients. 
For both types of carriers, a Poole-Frenkel-like behavior was observed with a negative field dependent mobility for almost all studied ratios.
Furthermore, we found double transients for both charge carrier types (see Fig.~\ref{fig:Fig1}): For $x=0.66$, we observe two extraction signals in the hole current transients. The same is true for the electron current transient for $x=0.8$ and $x=0.5$. We note that these transit times shift with the electric field, which is a clear indication that a field dependent charge carrier mobility instead of an artefact is measured.
\begin{figure}
	\includegraphics{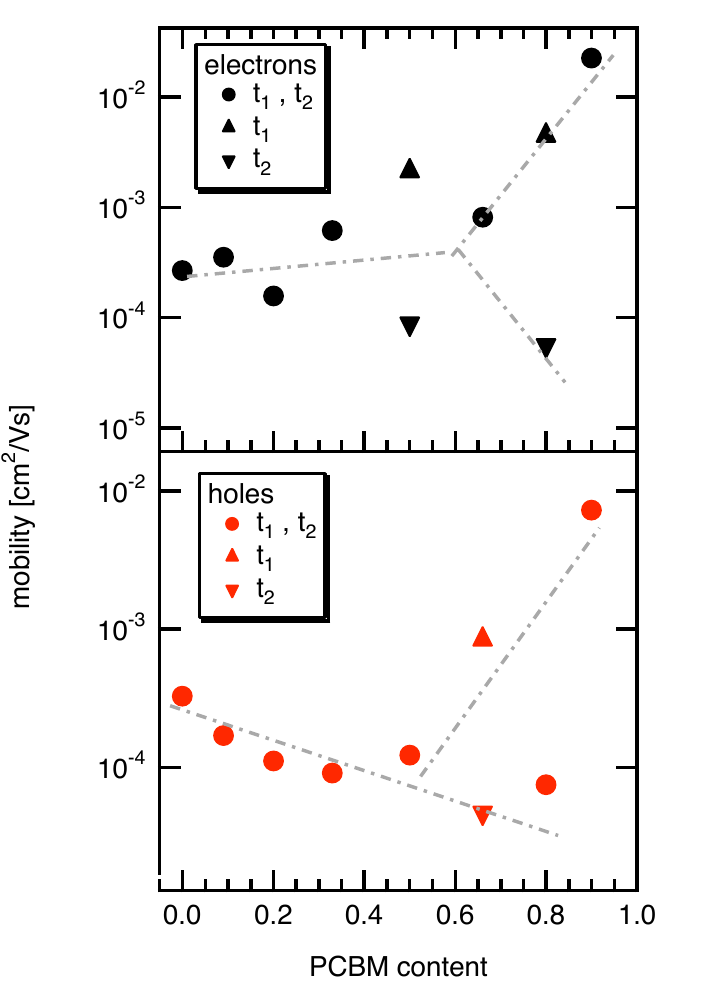}
	\caption{(Color online) Electron and hole mobilities in P3HT:PCBM blend films versus PCBM concentration at T=300K; mobility values were obtained by interpolating the Poole-Frenkel like field dependent mobility at an electric field of $2.5\cdot10^{7}$ V/m; for $x=0.8$ and $x=0.5$ a double transient for electrons and for $x=0.66$ for holes can be detected; circles denote transients with only one transit time and the triangles those with two transit times $t_1$ and $t_2$ observed. Guides to the eye are included (dashed lines).}
	\label{fig:Fig2}
\end{figure}
The electron and hole mobility for different PCBM content is summarized in Fig.~\ref{fig:Fig2}. We interpolated the mobility from the field dependent measurement at an external electric field of $2.5\cdot10^{7}$ V/m. A very similar dependence on the PCBM fraction was found by plotting mobilities at other electric fields.
We find a bipolar transport for the whole range of the studied P3HT:PCBM ratios. 
The electron mobility increases strongly with raising PCBM concentration above $50$\%. Below 50\%, the electron mobility remains almost constant or decreases slightly. The electron mobility rises up to $2.3\cdot10^{-2}$ cm$^2$/Vs for $x=0.9$, comparable to the mobility reported for PCBM in an insulating matrix~\cite{tuladhar2005}. In contrast, the hole mobility slightly decreases with increasing PCBM content in the blend for $x<0.25$, which might be related to a distortion of the semi-crystalline packing of P3HT. Below 75\% PCBM concentration the hole mobility varies little with adding PCBM content. We observe an increased hole mobility at 90\% PCBM content of about $7.3\cdot10^{-3}$ cm$^2$/Vs, which is comparable to the high electron mobility found in the same device.
In previously reported TOF measurements on MDMO-PPV:PCBM bulk heterojunction solar cells, the hole and the electron mobility increased with raising PCBM fraction~\cite{tuladhar2005}, which was explained by proposing a change in the morphology, in the sense that the added PCBM molecules straighten the PPV polymer chains~\cite{melzer2004}. However, the authors note that the high hole mobility might be due to holes hopping on PCBM, as they found a high hole mobility also in PCBM embedding in an insulating polystyrene matrix.\\

The mechanism responsible for double transients, signified by two different transit times, originates from slower charges being hindered in the extraction process by an energy barrier. Two potential explanations come to mind. First, the slow charge carriers become trapped in deep states during their hopping transport. Depending on the activation energy needed to leave these states, the charge transport of the trapped fraction of carriers is delayed as compared to the faster charge carrier package. 
These trap states can be caused e.g. by isolated P3HT molecules embedded in a percolated PCBM, where holes can be trapped, or PCBM molecules in an P3HT environment as electron traps. 
Second, the different mobilities can be due to electrons not only being transported on the PCBM, but --- once overcoming the energy barrier to hop onto the polymer phase --- also on the donor material, and vice versa.
However in case of trapping, we would expect the mobility to be lower than the one we calculate from the second transit time $t_2$.
Furthermore, we find from Monte Carlo simulations~\cite{deibel2007} in a gaussian density of states distribution without traps, that an increased phase separation in the blend leads to TOF transients with two clearly distinguishable transit times. In contrast, randomly distributed phases in the blend show only one single transit time. These results will be published elsewhere.
We note that double transients were also observed in liquid crystals being due to electronic and ionic contribution~\cite{iino2005a}.\\

In conclusion, we investigated the electron and hole mobility in bulk heterojunction solar cells with a variable P3HT:PCBM ratio, using the transient photoconductivity technique. We find a bipolar transport for all studied devices ranging from pure polymer to polymer:fullerene with 90\% PCBM content. In the PCBM concentration range $x<0.5$, the electron mobility remains nearly constant, whereas the hole mobility slightly decreases for $x<0.25$. For more than 50\% PCBM, both electron and hole mobility increases at 90\% PCBM concentration.
We observe two transit times in the electron transients for a PCBM content of $x=0.8$ and $x=0.5$ as well as double transient in the hole current transient for $x=0.66$. As a model, we propose that the slower carriers have to overcome a barrier before they can leave the sample, leading to longer extraction times as compared to faster charges. Although the origin for the observation of such double transients is unresolved yet, we devise two possible explanations: (a) decreased mobility due to trapping of carriers; (b) bipolar transport, with energetic activation barriers for electrons hopping on P3HT and holes on PCBM. For carrier trapping, we would in principle expect a lower mobility than experimentally observed.  
We find a high hole mobility on PCBM of approximately $7.3\cdot10^{-3}$ cm$^2$/Vs in the P3HT:PCBM 1:10 sample, which is comparable to the high electron mobility of $2.3\cdot10^{-2}$ cm$^2$/Vs found in the same device. Additionally, we observe ambipolar transport in pure P3HT sample with electron and hole mobilities in the range of $1.8\cdot10^{-4}$ cm$^2$/Vs.

\begin{acknowledgments}

A.B. thanks the German Federal Environmental Foundation (Deutsche Bundesstiftung Umwelt, DBU) for funding. V.D.'s work at the ZAE Bayern is financed by the Bavarian Ministry of Economic Affairs, Infrastructure, Transport and Technology. 

\end{acknowledgments}

\end{document}